# Toward sensitive document release with privacy guarantees

David Sánchez[a], Montserrat Batet[b], [1]

[a]*UNESCO Chair in Data Privacy, Department of Computer Science and Mathematics,*
*Universitat Rovira i Virgili, Avda. Països Catalans, 26, 43007 Tarragona (Spain)*
*david.sanchez@urv.cat*

[b]*Internet Interdisciplinary Institute (IN3), Universitat Oberta de Catalunya,*
*Parc Mediterrani de la Tecnologia, Av. Carl Friedrich Gauss, 5, 08860, Castelldefels, (Spain)*
*mbatetsa@uoc.edu*

**Abstract**

Privacy has become a serious concern for modern Information Societies. The sensitive nature of much of the data that are daily exchanged or released to untrusted parties requires that responsible organizations undertake appropriate privacy protection measures. Nowadays, much of these data are texts (e.g., emails, messages posted in social media, healthcare outcomes, etc.) that, because of their unstructured and semantic nature, constitute a challenge for automatic data protection methods. In fact, textual documents are usually protected manually, in a process known as document *redaction* or *sanitization*. To do so, human experts *identify* sensitive terms (i.e., terms that may reveal identities and/or confidential information) and *protect* them accordingly (e.g., via removal or, preferably, generalization). To relieve experts from this burdensome task, in a previous work we introduced the theoretical basis of *C-sanitization*, an inherently semantic privacy model that provides the basis to the development of automatic document redaction/sanitization algorithms and offers clear and a priori privacy guarantees on data protection; even though its potential benefits *C-sanitization* still presents some limitations when applied to practice (mainly regarding flexibility, efficiency and accuracy). In this paper, we propose a new more flexible model, named (*C*, *g*(*C*))-*sanitization*, which enables an intuitive configuration of the trade-off between the desired level of protection (i.e., controlled information disclosure) and the preservation of the utility of the protected data (i.e., amount of semantics to be preserved). Moreover, we also present a set of technical solutions and algorithms that provide an efficient and scalable implementation of the model and improve its practical accuracy, as we also illustrate through empirical experiments.

[1] Corresponding author. Address: Internet Interdisciplinary Institute (IN3), Universitat Oberta de Catalunya,
Av. Carl Friedrich Gauss, 5, 08860, Castelldefels. Spain. Tel.: +034 977 559657; E-mail: mbatetsa@uoc.edu

# 1. Introduction

Information Technologies have paved the way for global scale data sharing. Nowadays, companies, governments and subjects exchange and release large amounts of electronic data on daily basis. However, in many occasions, these data refer to personal features of individuals (e.g., identities, preferences, opinions, salaries, diagnoses, etc.), thus causing a serious privacy threat. To prevent this threat, appropriate data protection measures should be undertaken by responsible parties in order to fulfill with current legislations on data privacy [1, 2].

Because of the enormous amount of data to be managed and the burden and cost of manual data protection [3], many automated methods have been proposed in recent years under the umbrella of *Statistical Disclosure Control* (SDC) [4]. These methods aim at masking input data in a way that either *identity* or *confidential attribute* disclosure are minimized. The former deals with the protection of information that can re-identify an individual (e.g., a social security number or unique combinations of several attributes, such as the age, job and address), and it is usually referred to as *anonymization*, whereas the latter deals with the protection of *confidential* data (e.g., salaries or diagnosis). To do so, protection methods remove, distort or coarse input data while balancing the trade-off between privacy and data utility: the more exhaustive the data protection is, the higher the privacy but the less useful the protected data becomes as a result of the applied distortion, and vice-versa. In addition to data protection methods, the computer science community has proposed formal *privacy models* [5], within the area of *Privacy-Preserving Data Publishing* (PPDP) [6] and *Data Mining* (PPDM) [7, 8]. In comparison to the ad-hoc masking of SDC methods, in which the level of protection is empirically evaluated *a posteriori* for a specific dataset [5], privacy models attain a *predefined* notion of privacy and offer *a priori* privacy guarantees over the protected data (e.g., a probability of re-identification [9, 10]). This provides a clearer picture on the level of protection that is applied to the data, regardless the features or distribution of a specific dataset. Moreover, privacy models provide a *de facto* standard to develop privacy-preserving tools, which can be objectively compared by fixing the desired privacy level in advance.

So far, most privacy models and protection mechanisms have focused on structured statistical databases [11], which present a regular structure (i.e., records refer to individuals that are described by a set of usually uni-valued attributes) and mostly contain numerical data. Privacy models such the well-known *k-anonymity* notion relied on such regularities to define privacy guarantees: a data base is said to be *k-anonymous* if any record is indistinguishable with regard to the attributes that may identify an individual from, at least, *k-1* other records [9, 10].

However, many of the (sensitive) data that is exchanged in current data sharing scenarios is textual and unstructured (e.g., messages posted in social media, e-mails, medical reports, etc.). In comparison with structured databases, plain textual data protection entails additional challenges:

- Due to their lack of structure, we cannot pre-classify input data according to identifying and/or confidential attributes, as most data protection mechanisms do [11]; in fact, for plain text, any combination of textual terms of any cardinality may produce disclosure.
- In comparison with the usually numerical attributes found in structured databases, plain textual data cannot be compared and transformed by means of standard arithmetical operators. In fact, since textual documents are interpreted by data producers and consumers (and also potential attackers) according to the meaning of their contents, linguistic tools and semantic analyses are needed to properly protect them [12].

Because of the above challenges, the protection of plain textual documents has not received enough attention in the current literature [13-15]. As we discuss in the next section, most of the current methods and privacy models for textual data protection are naïve, unintuitive, require from a significant intervention of human experts and/or limit the protection to predefined types of textual entities.

## 1.1. Background on plain textual data protection

Traditionally, plain textual data protection has been performed manually, in a process by which several experts detect and mask terms that may disclose identities and/or confidential information, either directly (e.g., names, SS numbers, sensitive diseases, etc.) or by means of *semantic inferences* (e.g., treatments or drugs that may reveal sensitive diseases, readings that may suggest political preferences or habits that can be related to religion or sexual orientations) [16]. In this context, data semantics are crucial because they define the way by which humans (sanitizers, data analysts and also potential attackers) understand and manage textual data.

In general, plain textual data protection consists of two main tasks: i) identify textual terms that may disclose sensitive information according to a privacy criterion (e.g., names, addresses, authorship, personal features, etc.); and ii) mask these terms to minimize disclosure by means of an appropriate protection mechanism (e.g., removal, generalization, etc.). The community refers to the act of removing or blacking-out sensitive terms as *redaction*, whereas *sanitization* usually consists in coarsening them via generalization (e.g., *AIDS* can be replaced by a less detailed generalization such as *disease*) [3]. The latter approach, which we use in this paper, better preserves the utility of the output.

To relieve human experts from the burden of manual sanitization, the research community has proposed mechanisms to tackle specific data protection needs. On the one hand, we can find works that aim at inferring sensitive information, such as the authorship of a resource (e.g., documents, emails, source code, etc.) [17] or the profile of the author (e.g., gender) [18]; on the other hand, other works aim at preventing disclosure by masking the data that may disclose that authorship [19, 20]. In the healthcare context, we can find ad-hoc data protection approaches that focus on detecting *protected health information* (PHI, such as ages, e-mails, locations, dates or social security numbers) [21], which are data that, according to the HIPAA "Safe Harbor" rules, must be eliminated before releasing electronic healthcare records to third parties. Most of these application-specific approaches exploit the regularities of the lexico-syntactic regularities of the entities to be detected (e.g., use of capitalizations for proper names, structure of dates or e-mails, etc.) to define patterns or employ machine learning techniques such as trained classifiers. However, the applicability of these methods is limited to the use case they consider, and they do not offer robust guarantees against disclosure outside the entities in which they focus.

General-purpose privacy solutions for plain text are scarce and they only focus on the protection of sensitive terms, which are assumed to be manually identified beforehand. We can find two privacy models that reformulate the notion of *k*-anonymity for documents rather than data bases: *K*-safety [22] and *K*-confusability [23]. Both approaches assume the availability of a large and homogenous collection of documents, and require each sensitive entity mentioned in each document of the collection to be indistinguishable from, at least, K-1 other entities in the collection. To do so, terms are generalized (so that they become less diverse and, hence, indistinguishable) in groups of *K* documents. However, documents cannot be sanitized individually and, due to the need to generalize terms to a common abstraction, data semantics will be hampered if the contents of the collection are not perfectly homogenous.

In [15], a privacy model named *t*-plausibility that also relies on the generalization of manually identified sensitive terms was presented. A document is said to fulfill *t*-plausibility if, at least, *t* different *plausible* documents can be derived from the protected document by specializing sanitized entities; that is, the protected document generalizes, at least, *t* documents obtained by combining specializations of the sanitized terms. Even though this approach allows sanitizing documents individually, it is noted that setting the *t*-plausibility level is not intuitive and that one can hardly predict the results of a given *t*, because they would depend on the document size, the number of sensitive entities and the number of available generalizations and specializations.

To tackle the limitations of the above-described solutions, in [13] we presented an inherently semantic privacy model for textual data: *C-sanitization*. Its goal is to mimic and, hence, automatize the analysis of semantic inferences that human experts perform for document sanitization. Informally, the disclosure risk caused by semantic inferences is assessed by answering to this question: does a term or a combination of terms in a document to be released allow to univocally inferring and, thus, disclosing a sensitive entity defined in *C*? According to such vision, the privacy guarantees offered by the model state that a *C-sanitized* document should not contain any term that, individually or in aggregate, univocally reveals the semantics of the sensitive entities stated in *C*. In accordance with current privacy legislations, *C* may contain the entities that legal frameworks define as sensitive, such as religious and political topics or certain diseases [24]. For example, an *AIDS-sanitized* medical record should not contain terms that enable a univocal inference of AIDS, such as HIV or closely related symptoms or treatments.

In [13], *C-sanitization* is formalized according to the following elements: (1) D: the document to be protected. (2) *C*: the set of sensitive entities that should be protected from univocal disclosure in *D* (e.g., *C* could be a set of sensitive diseases or religious or political topics and *D* a medical record or a message to be posted in a social network). (3) *T*: whatever group of terms of any cardinality occurring in *D* that could be used by an attacker to unambiguously infer any of the sensitive entities in *C* (e.g., if *C* is a sensitive disease, *T* could be a synonym or a lexicalization, or a combination of treatments, drugs or symptoms that univocally refers to *C*). (4) *K*: the knowledge that potential attackers can exploit to perform the semantic inferences. The larger and the more complete the knowledge *K* is assumed to be, the stricter and the more realistic the assessment of disclosure risks will be and, hence, the more robust the privacy protection will be. *C-sanitization* relies on the evaluation of the disclosure risk that terms in *D* cause with regard to *C* according to the background knowledge *K*. Moreover, the privacy guarantees offered by *C-sanitization* ensure that univocal semantic inferences/disclosure of any of the entities in *C* are prevented. Formally, it is defined as follows.

**Definition 1**. (*C-sanitization*). Given an input document *D*, the background knowledge *K* and a set of sensitive entities *C* to be protected, *D'* is the *C-sanitized* version of *D* if *D'* does not contain any term *t* or group of terms *T* that, individually or in aggregate, univocally disclose *any* entity in *C* by exploiting *K*.

The enforcement of *C-sanitization* relies on the foundations of the Information Theory to assess and quantify the semantics to be protected (defined by *C*) and those disclosed by the terms appearing in the document to be protected, much like humans experts do [3]. The implementation of the *C-sanitization* can provide the following advantages over the above-described works: i) automatic detection of terms that may cause disclosure of sensitive data via semantic inferences, a task that has been identified as one of the most difficult and time-consuming for human experts [3, 16], ii) utility-preserving sanitization based on accurate term generalization, iii) intuitive definition of the a priori privacy guarantees by means of linguistic labels (i.e., the set *C* of entities to be protected), instead of the abstract numbers used in all the former privacy models, and iv) individual and independent protection of documents (rather than homogenous document collections), regardless their content or structure.

## 1.2. Contributions and plan of this paper

In spite of its potential benefits, *C-sanitization* still presents some limitations when applied to practical settings. In this paper, we tackle three main aspects. First, with *C-sanitization* the degree of protection is fixed: all the entities in *C* are protected in the same way according to a fixed criterion of strict non-univocal disclosure. This may be too rigid and even insufficient in scenarios in which a stricter protection is needed: not only the entities in *C* should not be univocally disclosed, but also the ambiguity of the inferences should be large, so that we avoid plausible (even though non-univocal) disclosure. To solve this issue, we propose a new privacy model (which we name (*C, g(C)*)-*sanitization*) that offers additional guarantees of disclosure limitation on top of those offered by the plain *C-sanitization*. The additional parameter of the (*C, g(C)*)-*sanitization* enables to seamlessly configure the trade-off between the additional protection and the preservation of semantics, in a similar way that the *k* or *t* parameters do for *k*-anonymity or *t*-plausibility, respectively; but, we use intuitive linguistic labels (rather than abstract numbers) that give the user a clearer idea of the expected degree of protection and of semantic preservation. In this regard, our goal is to improve the flexibility and adaptability of the model instantiation to heterogeneous scenarios and privacy/utility preservation needs without impairing the intuitiveness of its instantiation and of the privacy guarantees it offers.

Second, like any other model that tries to balance the trade-off between privacy protection and data utility preservation, the enforcement of (*C, g(C)*)-*sanitization* is NP-hard in its optimal form [13]. To render the implementation practical and scalable, in this paper we also propose several heuristics that carefully consider data semantics to guide the sanitization process. Moreover, we also propose a flexible greedy algorithm incorporating these heuristics and providing a practical and scalable implementation of the proposed model. The algorithm also provides parameters to configure its behavior towards maximizing its scalability or the protection accuracy.

Third, as it will be discussed in the fourth section, the enforcement of both *C-sanitization* and (*C, g(C)*)-*sanitization* relies on an accurate assessment of the informativeness of terms, which is used to quantify the semantics they disclose. Being able to perform such assessment in a generic way is not trivial [25, 26], and natural language-related problems (i.e., language ambiguity) may severely hamper its accuracy. To tackle this issue we also propose an accurate, scalable and generic mechanism to measure the informativeness of terms by using the Web as general-purpose corpora.

Finally, we illustrate the applicability and flexibility of (*C, g(C)*)-*sanitization* (in comparison with the former *C-sanitization*, which we use as evaluation baseline) by means of an empirical study. In this study, we also test the improvements related to protection accuracy and efficiency brought by the technical solutions we propose here with respect to the former work [13].

The rest of the paper is organized as follows. The second section presents the new (*C, g(C)*)-*sanitization*, which provides improved flexibility and configurability. The third section discusses the issues related to the practical enforcement of the model, proposes several heuristics to guide the protection process and presents a customizable and scalable algorithm. The fourth section discusses the issues related to the computation of term informativeness and proposes a generic solution that exploits the Web for that purpose. The fifth section reports and discusses the results of an empirical analysis of the model's implementation against the baseline work in [13]. The final section depicts the conclusions and presents some lines of future research.

## 2. A flexible privacy model for textual documents

In this section, we present a flexible privacy model for textual documents that allows configuring the trade-off between privacy protection (on top of the disclosure limitation guarantees of plain *C-sanitization*) and data utility preservation. As mentioned above, a *C-sanitized* document *D'* will offer the guarantee of a *non-univocal* disclosure (i.e., no unambiguous inference) of any entity in *C*. However, this guarantee could be too rigid in some scenarios. In practice, it is quite common to consider unacceptable the disclosure of a *significant* amount of the sensitive semantics because attackers may correctly infer the sensitive entities with a low ambiguity/high probability (even though not univocally). In these cases, we require of a mechanism to configure the trade-off between the additional degree of protection to be applied (i.e., a level of uncertainty in the semantic inferences larger than the strict non-univocal disclosure) and the preservation of data semantics allowed by such degree of protection.

The model we propose, named *(C, g(C))-sanitization*, allows configuring this trade-off on top of the privacy guarantees stated by *C-sanitization,* and without hampering the intuitiveness of the model instantiation. To do so, we define a (linguistic) parameter *g(C)* that allows to straightforwardly specify the maximum amount of allowed information/semantics disclosure of each entity *c* in *C*. To do so, we rely on the fact that the generalizations of an entity *c* disclose a strict subset of the semantics of *c*. According to this, we can lower the maximum level of semantic disclosure allowed for *c* (and, thus, force a stricter protection), by using an appropriate generalization *g(c)* as the threshold for risk assessment (instead of just *c*). Moreover, by defining a specific generalization for each *c,* we can independently and finely tune the allowed level of disclosure for each sensitive entity. This improves the flexibility of the model instantiation, which can be adapted to heterogeneous entities and privacy needs, as follows.

**Definition 2**. (*((C, g(C))-sanitization)* Given an input document *D,* the background knowledge *K*, an ordered set of sensitive entities *C* to be protected and an ordered set of their generalizations *g(C)*, we say that *D'* is the *(C, g(C))-sanitized* version of *D* if *D'* does not contain any term *t* or group of terms *T* that, individually or in aggregate, can disclose more semantics of *any* entity *c* in *C*, than those provided by their respective generalization *g(c)* in *g(C)* by exploiting *K*.

For example, if we apply an (*AIDS*, *chronic disorder*)-*sanitization* over a document, we are stating that the protected version will reveal an amount of semantics of *AIDS* that, at most, corresponds to those of its generalization, *chronic disorder*; that is, any conclusions resulting from a semantic inference more specific than that will be uncertain. On the contrary, an *AIDS-sanitized* document (i.e., according to Definition 1), even though would not univocally reveal the concrete disease, *AIDS*, may disclose more specific semantics that could enable to infer that the document is referring about a *disorder of the immune system*, a conclusion that may be risky in some scenarios. In general, the more abstract the generalizations *g(C)* used as thresholds are (e.g, *g(AIDS)=condition*), the more the ambiguity we add in the attacker's inferences and the less specific or certain his conclusions will be. Indeed, this lowers the actual disclosure risk at the cost of data semantics preservation, because a larger number of plausible solutions exists (e.g., in an (*AIDS, condition*)-*sanitized* document, conditions other than *AIDS* are as plausible as *AIDS*). Notice that, according to Definition 2, if several entities should be protected for a certain document, an ordered set of sensitive entities and their corresponding generalizations should be provided, such as ({*AIDS, HIV*}*,* {*Condition, Virus*}).

The disclosure assessment of *C-sanitization* is based on an information theoretical characterization of the semantics of terms. The underlying idea is that the semantics encompassed by a term can be quantified by the *amount of information* it provides, that is, its *Information Content* (IC), as it is widely accepted by the semantic community [25, 27]. By applying the notion of IC to each sensitive entity $c$ in $C$, it turns that $IC(c)$ is measuring the amount of *sensitive* information (of $c$) that should be protected because the disclosure of this *information* in the output document is what univocally reveals the semantics of $c$. Under the same premise, the amount of semantics of $c$ revealed by individual terms $t$ or groups of terms $T$ appearing in the document $D$ can be measured according to their overlap of information, that is, their *Point-wise Mutual Information* (PMI).

On the one hand, the IC of a textual term $t$ can be computed as the inverse of its probability of occurrence in corpora (which, in our case, represents the knowledge $K$ available to potential attackers).

$$IC(t) = -\log p(t) \tag{1}$$

On the other hand, the PMI between a term $t$ and a sensitive entity $c$ can be computed as the difference between the normalized probability of co-occurrence of the two entities, given their joint and marginal distributions in corpora [28]:

$$PMI(c;t) = \log \frac{p(c,t)}{p(c)p(t)} \tag{2}$$

Fig. 1 (left) shows how PMI measures the amount of information overlap between two entities.

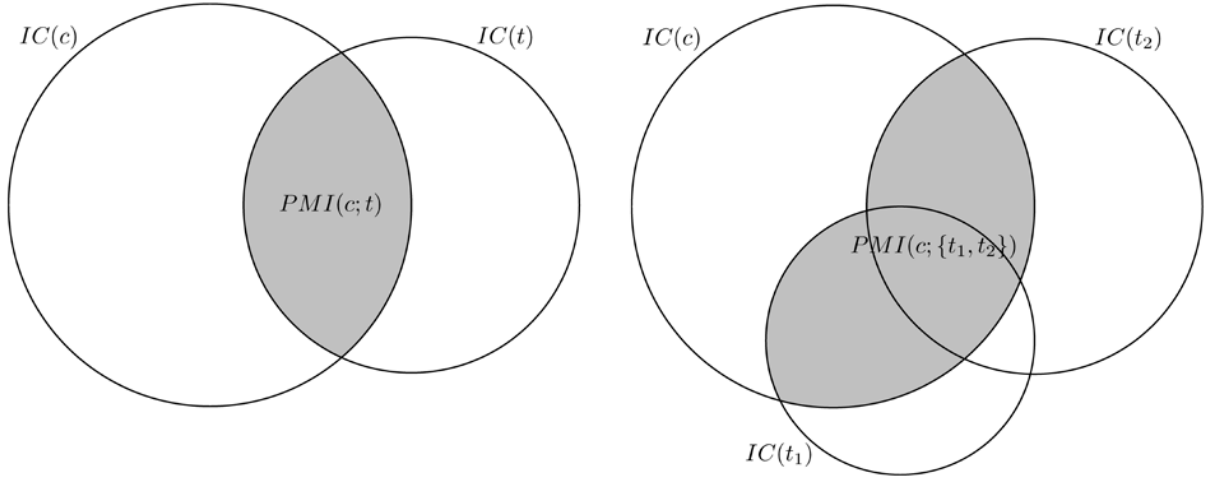

Fig. 1. Greyed area: on the left, amount of information/semantics that $t$ discloses from $c$ (and vice-versa); on the right: amount of semantics/information of $c$ disclosed by the co-occurrence of $t_1$ and $t_2$.

Likewise, the semantic disclosure of $c$ caused by the aggregation of a group of co-occurring terms $T=\{t_1,...,t_n\}$ can be computed as follows:

$$PMI(c;T) = \log \frac{p(c,t_1,...,t_n)}{p(c)p(t_1,...,t_n)} \qquad (3)$$

As shown in Fig. 1 (right), in this case, PMI is measuring the disclosure of $c$ as the *union* of the individual disclosures caused by each element $t_i$ in the group $T=\{t_1, t_2\}$.

Numerically, PMI is maximum if, in the underlying corpora that represents the knowledge $K$, a single $t$ or a combination $T$ always co-occurs with $c$, thus resulting in $PMI(c;t)=IC(c)$ and $PMI(c;T)=IC(c)$, respectively. This states that $c$ is completely disclosed by $t$ or $T$ because the semantics of the former can be univocally inferred from the latter (i.e., there is no ambiguity in the semantic inference).

Thus, to satisfy Definition 1, those individual terms $t$ or groups of terms $T$ in $D$ whose PMI with regard to each $c$ in $C$ is equal to the IC($c$), should be sanitized or redacted (i.e., generalized or removed) from the output document $D'$.

By relying on the information theoretic characterization of data semantics depicted above, we propose enforcing (*C, g(C))-sanitization* (Definition 2) as follows. First, instead of using *IC(c)* as the *threshold* stating the maximum allowed disclosure during the assessment of risks, we use *IC(g(c))*; in this way, the maximum amount of information/semantics of *c* that is allowed to be disclosed in the protected document are lowered until those of *g(c)*, which is strictly less informative than *c*. This is formalized in Definition 3.

**Definition 3**. (*Information Theoretic* (*C, g(C)*)-*sanitization*). Given an input document *D,* the corpora that represents the knowledge *K*, an ordered set of sensitive entities *C* to be protected and an ordered set of their generalizations *g(C)*, we say that *D'* is the (*C, g(C)*)-*sanitized* version of *D* if, for all *c* in *C, D'* does not contain any term *t* or group of terms *T* so that, according to corpora, $PMI(c;t) > IC(g(c))$ or $PMI(c;T) > IC(g(c))$, respectively.

Graphically, as shown in Fig. 2 (right), the use of a generalization *g(c)* as disclosure threshold for *c* (boldface circled area) lowers the amount of information/semantics that terms *t* or groups of terms *T* can disclose about *c* (greyed area). Compared to the basic *C-sanitization* (Fig. 2 (left), in which *IC(c)* acts as the threshold and for which *t* is not risky), (*C,g(C)*)-*sanitization* forces the system to implement a stricter sanitization (see Fig.2 (right), in which *IC(g(c))* is the threshold and for which *t* is risky); this will provide a better protection at the cost of the preservation of data semantics.

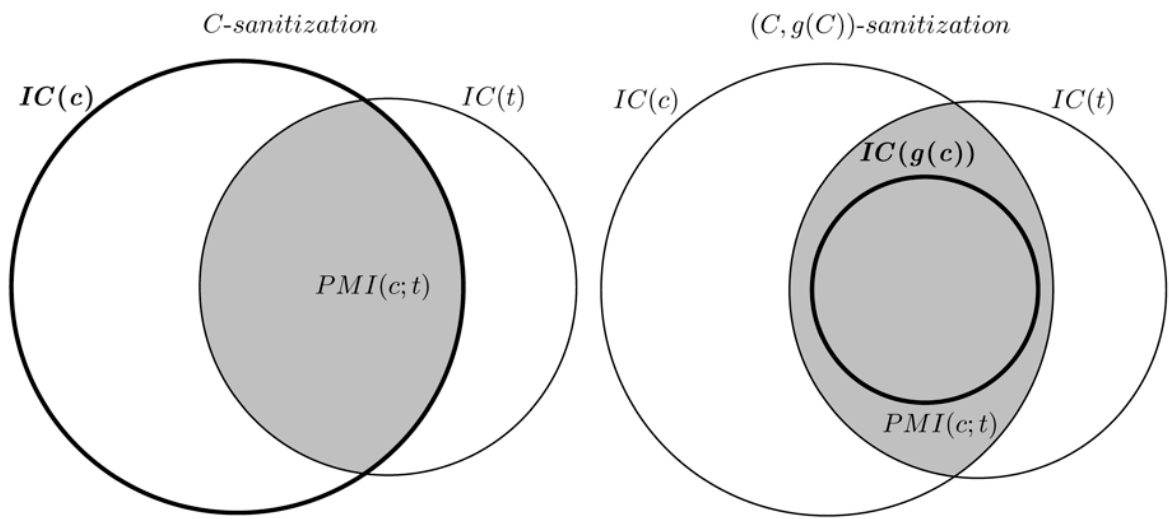

Figure 2. Left: *C-sanitization*; Right: (*C,g(C)*)-*sanitization*. Boldface circled areas represent the thresholds used by each model for assessing disclosure risks.

The *g(c)* parameter in the proposed model allows configuring the inherent trade-off between privacy protection and data utility preservation, similarly to what the numerical parameters of other models (e.g., *k*-anonymity, *t*-plausibility) do. For example, the larger the *k* we specify when instantiating the *k*-anonymity model, the more homogenous (i.e., indistinguishable) the protected data become and, thus, the better the protection is but the lower the utility will be. In (*C,g(C)*)-*sanitization*, the more abstract the generalizations *g*(*C*) are, the less informative but the more protected *D'* becomes. Even though the possibility of balancing this trade-off is common to most privacy models available in the literature [11], all of them rely on abstract numerical parameters (*k*-anonymity, *l*-diversity, *t*-closeness, *ε*-differential privacy, *t*-plausibility, *K*-confusability, *K*-safety) whose practical influence in the protected output is, in most cases, difficult to understand and even more difficult to predict [13, 15]. On the contrary, the use of generalizations as thresholds of disclosure in our model is very intuitive because it provides a clear understanding on the amount of semantics that external entities (whether they are readers, data analysts or attackers) can learn of each *c* in the protected document. Moreover, as discussed above and contrary to numerically-oriented privacy models, these linguistic parameters enable a seamless adaptation of the model instantiation to current legislations on data privacy, whose rules about the topics that should be protected and up to which degree are also expressed linguistically; for example, *locations* more specific than *counties* should be protected according to the HIPAA [1], information that could reveal the specific *race* or *religion* of an individual is sensitive according to the EU Data Protection Regulation, etc. This greatly facilitates the model instantiation and, as far as we know, provides the first privacy model by which practitioners can directly enforce the guidelines stated in current legal frameworks.

## 3. Towards scalable and utility-preserving sanitization of risky terms

In order to be utility preserving, document sanitization should protect risky terms by replacing them by generalizations, rather than just removing or blacking them out. Generalizing risky terms *t* (e.g., *AIDS*) by privacy-preserving generalizations *g*(*t*) (e.g., *disease*), which are less specific (i.e., $IC(disease)<IC(AIDS)$), will decrease the level of disclosure of *c* while still retaining a *subset* of the semantics of *t* (i.e., we still know that *t* is a *disease*). Thus, in our approach, once a term *t* is found to be risky according to the model instantiation, it is replaced with an appropriate generalization *g*(*t*) retrieved from a Knowledge Base (KB) that fulfills the privacy guarantee as defined in Definition 3; considering that the sanitized document *D'* must not contain any term *t* so that $PMI(c;g(t)) > IC(g(c))$, we replace terms by generalization so that $PMI(c;g(t)) \leq IC(g(c))$, where $PMI(c;g(t))$ represents the information retained of *t* (and disclosed of *c*) when *t* is replaced by a generalization *g*(*t*).

Fig.3 illustrates this process: a term *t*, which discloses more information about *c* (whole greyed area) than that allowed by the threshold *g*(*c*) (boldface circle) is replaced by *g*(*t*) (dashed circle), which lowers the disclosed information low enough below the threshold but still retains some semantics (dark greyed area).

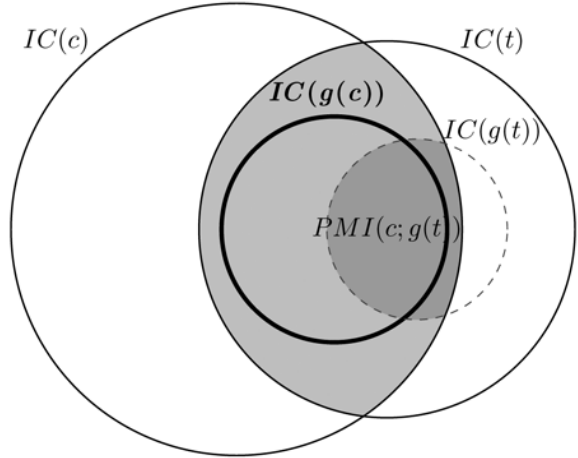

Fig. 3. *(C,g(C))-sanitization* of *t* via utility-preserving generalization: *t* is replaced by *g(t)*.

From the perspective of the preservation of data utility, the optimal generalization *g(t)* replacing a term *t* should retain the maximum semantics of *t* as possible, while fulfilling the guarantees of the model instantiation; in other words, *g(t)* should be the generalization, from those available in the KB, with the highest $IC(g(t))$ that fulfills $PMI(c;g(t)) \leq IC(g(c))$. As introduced above, like any other model that tries to balance the trade-off between privacy protection and data utility (e.g., *k*-anonymity, *t*-plausibility) selecting these generalizations in an optimal way (with respect to data utility preservation) is NP-hard; thus, it could compromise the applicability of the model. In this section, we discuss this issue and propose several technical solutions to make the model implementation scalable.

On the one hand, it is important to note that the selected generalization *g(t)* (from those available in the KB) of a certain *t* should fulfill the privacy criterion for *all* of the sensitive entities $c_j$ in *C*. Formally:

$$g(t) = \underset{\forall g_i(t) \in KB | (PMI(c_j; g_i(t)) \leq IC(g(c_j)), \forall c_j \in C)}{\arg\max} (IC(g_i(t))) \quad (4)$$

On the other hand, in the most general case of groups of terms *T* of any cardinality, the optimal sanitization (i.e., the combination of generalizations of each $t_i$ in *T* that, in aggregate, retains the maximum amount of semantics while fulfilling the privacy criterion) is certainly NP-hard because the order in which terms in *D* are evaluated influences the order in which the groups of terms *T* are analyzed and sanitized, if needed. Hence, an optimal utility-preserving sanitization requires evaluating *all* possible combinations of *T* in *D* of any cardinality and any possible generalization of each $t_i$ in *T*, and picking up the combination of *T* that fulfills the privacy criterion while optimizing the preservation of semantics.

To render the sanitization process practical, we propose the following efficient greedy algorithm that relies on several utility-preserving heuristics to provide a scalable implementation of the model.

**Algorithm 1.**

```
Input:  D      //the input document
        KBs    //the knowledge bases used to retrieve generalizations
        C      //the ordered set of entities to be protected
        g_C    //the ordered set of generalizations of the entities to be protected
        MAX    //maximum cardinality of the combinations of terms (A2)
Output: D'     //the sanitized document

1  D'=D;
2  for each (d_i ⊆ D) do //for each context d_i defined in the document D (A1)
3    n=1; //cardinality of the combination of terms to evaluate (H1)
4    Term_set_i=getSortedTerms(d_i); //terms in the context d_i sorted by their IC (H2)
5    while (n≤|Term_set_i| and n<MAX) do //create combinations up to MAX cardinality(A2)
6      Comb_in=getComb(Term_set_i, n); //ordered set of combinations with cardinality n
7      while (not(empty(Comb_in))) do //evaluate each combination
8        T_j=first(Comb_in);
9        remove(T_j, Comb_in);
10       risky =false;
11       c_k=first(C); //obtain the first sensitive entity to evaluate
12       g_c_k=first(g_C); //get the corresponding generalization
13       while (not(risky) and c_k≠null) do //evaluate all sensitive entities
14         if (PMI(c_k;T_j)>IC(g_c_k)) then //privacy criterion according to Def.4
15           risky =true;
16         else
17           c_k=next(C); //get the next sensitive entity
18           g_c_k=next(g_C); //get the corresponding generalization
19         end if
20       end while
21       if (risky) then //if the combination was risky
22         Gen_set_j=getSortedGen(T_j, KBs); //ordered sets of generalizations of T_j (H3)
23         g_T_j=first(Gen_set_j);
             //check if the generalization set g_T_j fulfills Def. 4 for all c_k in C
24         while (not(PMI(c_k;g_T_j)<(IC(g_c_k)) ∀ c_k ϵ C)) do
25           g_T_j=next(Gen_set_j);
26         end while
27         sanitize(T_j, g_T_j, D'); //replace terms in T_j by generalizations in g_T_j
```

```
28              replaceGenTerms(T_j, g_T_j, Term_set_i); //also in the set of terms to evaluate
29          end if
30      end while
31      n++; //evaluate longer combinations of terms
32   end while
33 end for
34 return D';
```

Three heuristics are proposed to guide the protection process towards maximizing the preservation of data semantics:

- *H1*: evaluate first the groups of terms $T$ with lowest cardinality $|T|$. The idea is to avoid starting the analysis with large sets of terms because, in case of disclosure of any $c$ in $C$, we cannot discern which of the terms $t_i$ in $T$ are indeed causing that disclosure (because of their strong relationship with $c$) from those that are irrelevant (and that could be even removed from $T$ without significantly altering the disclosure assessment). Thus, to avoid unnecessary sanitizations of non-related or slightly related terms when combined with highly related ones, we start the analysis with individual terms (i.e., $|T|$=1, line 3 in Algorithm 1). In case of disclosure, these individual terms will be sanitized and, from that moment, they will be replaced by the appropriate generalizations in further iterations of the analysis (line 28 in Algorithm 1). Only once all individual terms have been evaluated for disclosure (and eventually replaced by generalizations), the process continues with groups of two terms, and so on (line 32 in Algorithm 1).
- *H2*: at each iteration of the analysis, evaluate first those terms with the highest informativeness (line 4 in Algorithm 1). Terms $t$ with high $IC(t)$ reveal more information; thus, if $t$ is semantically related with the entities $C$ to be protected (most textual terms are in fact related up to some degree [14]), those terms with a higher informativeness are more likely to disclose a larger amount of the $C$'s semantics. In this manner, the most potentially risky terms and the combinations in which they are involved are evaluated (and sanitized) first (line 6 in Algorithm 1). With this, we contribute to minimize the amount of unnecessary sanitizations, because we avoid combining highly informative (and thus potentially risky) terms with those with low informativeness, which may not need to be protected.
- *H3*: once a combination of terms $T$ is found to be risky according to the privacy criterion, the sanitization process is performed by: i) picking all the possible generalizations $g(t_i)$ for all the terms $t_i$ in $T$, ii) generating all the possible combinations of generalizations, iii) sorting these combinations according to the aggregated informativeness of their elements (line 22 in Algorithm

1) and, according to such order, iv) selecting the first one that fulfills the privacy criterion (line 24 in Algorithm 1).

We can see that the combination of these heuristics in the greedy algorithm guides the protection towards minimizing the number of unnecessary sanitizations. Moreover, in order to make the practical enforcement more scalable (especially for large documents) two additional parameters have been also incorporated into the algorithm:

- *A1*: allows splitting the input document *D* in several textual contexts $d_i \subseteq D$ (line 2 in Algorithm 1) so that the analysis of groups of terms is limited to the words framed in a subset of the document and, thus, a smaller amount of term combinations shall be considered. This relies on the assumption that, in general, the strength of the semantic relationships between terms co-occurring in a document tends to decrease as they are more distant in the text [29], as it does the disclosure risk. This approach is usually considered by other sanitization mechanisms available in the literature, which define textual contexts as paragraphs, sentences or just adjacent words [15].
- *A2*: limits the maximum cardinality of the groups of terms. As above, with this action we achieve analyzing a smaller number of combinations (line 5 in Algorithm 1). In the most extreme case, only individual terms may be considered, thus solving the problem in linear time with respect to the number of terms in the document. However, this action may hamper the accuracy of the disclosure assessment because larger combinations of terms will be omitted. To compensate this issue, more abstract generalizations *g(C)* may be defined as sanitization thresholds, so that a stricter (but more scalable) sanitization is applied in practice.

## 4. Corpora selection and probability calculation

The information theoretic enforcement of *(C, g(C))-sanitization* extensively relies on the probabilities of (co-)occurrence of terms (eqs. (1)(2)(3)) to measure term semantics. The accuracy of the probability calculation is thus crucial to ensure the consistency of the model enforcement; however, this is also a challenging task, because language ambiguity may hamper this calculation [26, 30]. On the other hand, the selection of the corpus from which to compute probabilities is also important to make *K* (i.e., the knowledge we assume is available for attackers) a faithful representation of the semantics as they are understood and used by humans (i.e., sanitizers, data analyst and potential attackers). In this section, we discuss these two important issues and propose a general approach to accurately compute probabilities.

The first author proposing the use of IC to quantify the semantics of concepts was Resnik [25], in a work that seminally inspired many other semantic researchers [31-35]. Probabilities were computed from tagged textual corpora, in which term appearances were manually associated to their conceptualizations (i.e., meanings) that were modeled in a reference knowledge base; for example, the occurrence of the word "virus" referring to a malicious computer program was associated to the concept "(computer) virus", whereas the occurrence of the word "virus" referring to an infectious microorganism was associated to the concept "(biological) virus". In this manner concept occurrences were properly disambiguated and, thus, their probabilities were not affected by the potential ambiguity (i.e., synonymy or polysemy) inherent to the terms (and their lexicalizations) used to refer to the concepts. Moreover, the probability calculation was made consistent with the taxonomic structure of the conceptualizations used to perform the annotation (e.g., *(biological) virus* → *microorganism* → *organism*): the probability of a concept *c* (e.g., *microorganism*) considers all the explicit appearances of the lexicalizations of *c* (i.e., synonyms or acronyms, such as *micro-organism*) and also all of their specializations (i.e., hyponyms, such as *virus*, *bacteria*, *pathogen*, etc.). Formally:

$$p(c) = \frac{\sum_{c_h \in hyponyms(c)} occurrences(c_h)}{N}, \qquad (5)$$

where *hyponyms*(*c*) includes the concept *c* and all of its specializations, and *N* is the total number of occurrences of concepts in the corpus.

With this approach, it is ensured that the probability of concepts monotonically increase as one moves up in the taxonomy of concepts (e.g., $p((biological)\ virus) \leq p(microorganism) \leq p(organism)$) which, in turn, ensures that the IC of a concept is lower than that of its hyponyms (e.g., $IC((biological)\ virus) \geq IC(microorganism) \geq IC(organism)$) [25]. This provides consistency to the IC as an assessor of the semantics of concepts because, as stated in section 3, generalizations must provide a strict subset of the semantics of their specializations. In fact, we recall that the sanitization implemented by our model extensively relies on the fact that, by replacing sensitive terms by appropriate generalizations, the amount of semantics disclosed is effectively reduced in order to fulfill the privacy model.

Unfortunately, since tagged corpora are commonly annotated manually, their size and coverage are limited, especially for domain-specific terms (e.g., technical or scientific terminology), named-entities (e.g., organization names) or newly minted terms (e.g., a new electronic device), which are the usual targets of document sanitization due to their high specificity/informativeness. Thus, the probability calculation may be hampered by data sparseness, which is obviously more prone to appear when computing the *co*-occurrences of several terms in which the PMI calculation relies.

To tackle these limitations, it is possible to use massive raw electronic corpora instead of tagged text. Contrary to tagged corpora, electronic raw resources are largely accessible in the Web and cover most domains of knowledge. In fact, the corpus of resources offered by the Web is so large and heterogeneous that it is said to be a faithful representation of the information distribution at a social scale [36]; this argument has been supported by recent works focusing on privacy-protection [14, 37], which considered the Web as a realistic proxy for social knowledge. In this respect, the Web provides a good representation of the background knowledge *K* that attackers may exploit to infer sensitive data [37] and, thus, we propose to use it for probability calculations; that is, we assume that a perfectly knowledgeable attacker is such that has the whole knowledge provided by the Web available. Moreover, within the Web context, occurrence and co-occurrence probabilities can be efficiently computed by querying terms in a publicly available Web Search Engine (WSE) and retrieving the resulting page count [38]; specifically, the IC of an entity and the PMI of a pair can be computed from the Web information distribution as follows:

$$IC(c) \cong -\log_2 \frac{page\_count("c")}{W} \qquad (6)$$

$$PMI(c;t) \cong \log_2 \frac{\frac{page\_count("c" \text{ AND } "t")}{W}}{\frac{page\_count("c")}{W} \times \frac{page\_count("t")}{W}} \qquad (7)$$

where *W* is the number of web resources indexed by the web search engine.

Likewise, to evaluate the disclosure that groups of terms *T* (where $T=\{t_1,...,t_n\}$ without ordering) cause with regard to a sensitive entity *c*, we can compute their PMI as follows:

$$PMI(c;T) = \log \frac{p(c,t_1,...,t_n)}{p(c)p(t_1,...,t_n)} \cong \log_2 \frac{\frac{page\_count("c" \text{ AND } "t_1" \text{ AND}...\text{AND} "t_n")}{W}}{\frac{page\_count("c")}{W} \times \frac{page\_count("t_1")}{W} \times ... \times \frac{page\_count("t_n")}{W}} \qquad (8)$$

A limitation of this calculation is the fact that probabilities computed from the WSE page count can be severely affected by language ambiguity [30]. When the words used to refer to a concept are polysemic (e.g., *virus*), the probability computed from the web page count will overestimate the actual probability of the underling concept (e.g., (*biological*) *virus*), because the count includes *all* the appearances of the word in the Web, regardless of the concept (i.e., meaning) to which they refer to; in this case, the concept will be considered *less* informative than what it truly is. Likewise, if a concept can be referred by means of different synonyms (e.g., the terms *HIV* and *human immunodeficiency virus* refer to the same disease) or entities referred in a discourse are not explicitly included in text due to ellipsis, the resulting probabilities will be lower and, thus, the informativeness will be overestimated. Even more important, and also related to language ambiguity, the monotonicity of the IC calculation with regard to the taxonomic subsumption will be hardly ensured. For example, querying *HIV* in Google provides 55 million results, whereas querying *microorganism*, which is a generalization of *HIV*, produces roughly 1.5 millions because many appearances of *HIV* do not explicitly mention *microorganism* due to ellipsis and synonymy. Taking these values independently would result in *HIV* having much lower informativeness than *microorganism*, which is inconsistent with respect to their taxonomic relationship (*HIV* is a specialization of *microorganism*). If not solved, this inconsistency will seriously hamper the applicability of our model, in which the informativeness of the generalizations (e.g., *retrovirus*→ *virus* → *microorganism*) is used as thresholds to configure the allowed disclosure of the semantics of the sensitive entities (e.g., *HIV*).

To minimize the problems related to language ambiguity and to ensure the monotonicity of the IC calculation in which our disclosure assessment relies, in the following we propose a mechanism to contextualize WSE queries of sensitive entities within the scope of the generalizations defined in the (*C, g(C)*)-*sanitization* instantiation. The main idea is that, by forcing the co-occurrence of a term referring to an entity (e.g., *virus*, as an infectious microorganism) and the generalization that is adequate to the meaning of the entity (e.g., *microorganism*), the effect of ambiguity in the resulting page count is minimized while the monotonicity of the IC calculation between the specialization and the generalization is fulfilled. On the one hand, polysemy is minimized because word occurrences rarely refer to different senses within the same document: if *virus* and *microorganism* co-occur in a document it is very unlikely that the former refers to a *computer program*. Likewise, since only explicit co-occurrences of the term and its generalization are considered, the potential ellipses of the latter are omitted from the probability assessment, thus fulfilling the monotonicity of the IC calculation (i.e., *page_count*(*"virus" AND "microorganism"*) < *page_count*(*"microorganism"*) and, thus, *IC*(*virus & microorganism*) > *IC*(*microorganism*)). The only drawback is the fact that the explicit contextualization of term occurrences will constraint the size of the sample considered in the calculation of probabilities (i.e., all the appearances of a (*biological*) *virus* alone are omitted). However, the size and redundancy of the Web helps to minimize the effect of this handicap, which, in any case, is preferable to the negative influence of language ambiguity and the lack of monotonicity [26].

In practice, to contextualize the page count resulting from the queries performed to the WSE, the appropriate generalization will be attached to the term to be queried by using a logic operator supported by the WSE, such as AND or +. This contextualization is applied to all the queries evaluating the disclosure risk of the sensitive entity *c*, so that the PMI calculation is made numerically coherent with the IC of the generalization *g(c)* that acts as threshold for the (*C, g(C))-sanitization* instantiation. Indeed, the generalization *g(c)*, which is picked up by the user of the model and that would correspond to the appropriate meaning of *c*, will implicitly disambiguate the occurrences of *c*. In this manner, only the occurrences of *c* that correspond to hyponyms of *g(c)*, which are the appropriate ones to measure the disclosure, will be considered in the calculation. Formally, we propose computing the PMI between a sensitive entity *c* and a term *t*, contextualized by the generalization *g(c)* defined in (*C, g(C))-sanitization*, which we denote as $PMI_{g(c)}(c;t)$, as follows:

$$PMI_{g(c)}(c;t) = \log_2 \frac{\frac{page\_count("c"\ AND\ "g(c)"\ AND\ "t")}{W}}{\frac{page\_count("c"\ AND\ "g(c)")}{W} \times \frac{page\_count("t")}{W}} \quad (9)$$

We apply this contextualized calculation to the disclosure risk assessment of (*C, g(C))-sanitization* (Definition 3), so that terms *t* or groups of terms *T* are risky if $PMI_{g(c)}(c;t) > IC(g(c))$ or $PMI_{g(c)}(c;T) > IC(g(c))$ for any *c* in *C*. Likewise, in case of disclosure and according to eq. (4), we replace *t* by the most informative generalization *g(t)* that fulfills $PMI_{g(c)}(c;g(t)) < IC(g(c))$ for all *c* in *C*.

Probability contextualization ensures the semantic and, thus, numerical consistency of the IC/PMI calculus in which the model enforcement relies. As a consequence of this, we also enable that any generalization of *c* could be used as threshold for instantiating (*C, g(C))-sanitization*, because the contextualization fulfills $PMI_{g(c)}(c; g(c)) = IC(g(c))$, which is coherent with the notion of taxonomic subsumption; that is, *g(c)* is completely disclosed by *c* because the semantics of the former can be univocally inferred from the latter.

## 5. Empirical analysis

In this section, we report and discuss a set of empirical results that show the suitability and benefits of (*C, g(C))-sanitization* and the technical solution presented above (i.e., the heuristic greedy algorithm and the web-based contextualized probabilities) over the basic *C-sanitization* model, which we use as baseline. To evaluate the differences between both approaches (in terms of privacy protection accuracy and runtime), we configured the following set of scenarios by enabling or disabling the following features proposed in the current work:

- *Model parameterization*. As proposed in section 2, the (*C, g(C)*)-*sanitization* instantiation can be tailored to specific privacy/utility needs by using the parameter *g(c)* (i.e., a generalization *g(c)* for each entity *c* to be protected) that states the maximum level of disclosure. To analyze the benefits of this parameter in the protected outcome, we defined several generalizations for the different entities considered in the evaluation and compared the results they obtained against those obtained against the (non-parameterized) *C-sanitization* [13].
- *Analysis of groups of terms*. As discussed above, the evaluation of combinations of terms of unbounded cardinality can be costly when dealing with large contexts and/or documents, because of the number of possible combinations to analyze. In our proposal, the cardinality of the groups of terms to be analyzed can be limited by means of the MAX parameter incorporated into Algorithm 1 (section 3), which allows reducing the number of elements to analyze. To evaluate the influence of this parameter both in the sanitized output and in the run time of the sanitization process, we performed different executions by fixing its value to 1 and 2. Due to the tight discourses of the evaluated documents (see details below), the whole text has been considered as a unique context.
- *Contextualization of probabilities*. In section 4, we discussed the issues related to the calculation of concept probabilities from the page count of a web search engine, and we proposed a way to minimize the effects of language ambiguity while ensuring that probabilities are coherent with the notion of taxonomic subsumption. To measure the benefits of this calculation methodology, we evaluated the sanitized output resulting from the standard probabilities (as done in [13]) and from the contextualized version we propose here. In all cases, we used the Bing (http://www.bing.com/) web search engine.

Our experimental case study is based on the privacy requirements stated in U.S. federal laws on medical data privacy [24], which mandate hospitals and healthcare organizations to protect medical concepts that are considered confidential before releasing patient records to, for example, insurance companies, in response to Worker's Compensation or Motor Vehicle Accident claims, or a judge, in case of malpractice litigation [3]. In particular, all references to Sexually Transmitted Diseases (STDs) or HIV status should be redacted or sanitized. To do so, terms explicitly referring to these diseases and those semantically related ones such as drugs, treatments or symptoms should be identified and protected [3].

Coherently with the healthcare scope, we used the Wikipedia articles describing the sensitive entities (*STD* and *HIV*) as the input documents to be protected. Wikipedia articles are commonly used by researchers on document sanitization [13, 14, 37, 39-42] because they are considered authoritative sources of information and also because of their semantically tight discourses, which configure a specially challenging scenario for data protection.

We have instantiated the (*C, g(C)*)-*sanitization* parameters in accordance with the semantics of the entities to be protected and according to the knowledge modeled in a standard knowledge base (WordNet [43]), which has been also used as the KB to retrieve the generalizations used to replace (sanitize) risky terms. We used *Virus* and *Infection* as the generalizations of the two senses of *HIV* (an infectious agent and an infectious disease) and *Disease* as the generalization of *STD*. Moreover, we also considered the case in which no generalizations are stated, as expressed in Definition 1 [13] .

The evaluation of the sanitized output has been carried out by comparing the former with the sanitization performed by a human expert that our proposal tries to mimic. To do so, the expert was requested to manually remove terms or groups of terms that, individually or in aggregate, may disclose the entity to be protected. Hereinafter, we refer as *H* to this set of manually removed terms, and as *S* to the set of terms detected by our proposal. The evaluation is performed by comparing the sets *H* and *S* according to the standard measures of *precision*, *recall* and *F-measure*.

*Precision* (eq. 4) quantifies the percentage of terms identified as sensitive by our proposal (*S*), which have been removed by the human expert (*H*). The higher the precision is, the better the utility of the output will be, because we are incurring in a lower number of non-necessary sanitizations.

$$Precision = \frac{|S \cap H|}{|S|} \times 100 \qquad (10)$$

*Recall* (eq. 5) quantifies the percentage of correctly sanitized terms from the total number of terms identified by the human expert. The higher the recall is, the better the protection will be.

$$Recall = \frac{|S \cap H|}{|H|} \times 100 \qquad (11)$$

*F-measure* (eq. 6) quantifies the harmonic mean of precision and recall and, thus, summarizes the accuracy of the protection process with respect to the human criterion.

$$F - measure = \frac{2 \times Recall \times Precision}{Recall + Precision} \qquad (12)$$

## 5.1. Results and discussion

The results obtained for the *C-sanitization* and (*C, g(C)*)-*sanitization*, the generalizations of each entity (i.e., *virus* and *infection* for *HIV* and *disease* for *STD*) and the different implementation parameters discussed earlier (i.e., MAX cardinality of groups of 1 or 2 and standard or contextualized probability calculations) are evaluated in Tables 1 and 2 for *HIV* and *STD*, respectively.

Table 1. Precision, recall and F-measure for different model instantiations (the first row corresponds to *C-sanitization* and the second and third to (*C, g(C)*)-*sanitization*) and implementation parameters for the *HIV* document.

| Model instantiation | Implementation parameters | Precision | Recall | F-measure |
| --- | --- | --- | --- | --- |
| *HIV*-sanitization | *MAX*=1; Standard probability | 100% | 23.8% | 38.4% |
| | *MAX*=1; Contextualized probability | N/A | N/A | N/A |
| | *MAX*=2; Standard probability | 65.5% | 90,5% | 76% |
| | *MAX*=2; Contextualized probability | N/A | N/A | N/A |
| (*HIV, Virus*)-sanitization | *MAX*=1; Standard probability | 88.9% | 38.1% | 53.3% |
| | *MAX*=1; Contextualized probability | 84.2% | 76.2% | 80% |
| | *MAX*=2; Standard probability | 61.3% | 100% | 76% |
| | *MAX*=2; Contextualized probability | 67.7% | 100% | 80.7% |
| (*HIV, Infection*)-sanitization | *MAX*=1; Standard probability | 100% | 33.3% | 50% |
| | *MAX*=1; Contextualized probability | 74% | 95.2% | 83.3% |
| | *MAX*=2; Standard probability | 53.8% | 100% | 70% |
| | *MAX*=2; Contextualized probability | 58.3% | 100% | 73.7% |

Table 2. Precision, recall and F-measure for different model instantiations (the first row corresponds to *C-sanitization* and the second to (*C, g(C)*)-*sanitization*) and implementation parameters for the STD document.

| Model instantiation | Implementation parameters | Precision | Recall | F-measure |
|---|---|---|---|---|
| *STD*-sanitization | *MAX*=1; Standard probability | 87.5% | 18.9% | 31.1% |
|  | *MAX*=1; Contextualized probability | N/A | N/A | N/A |
|  | *MAX*=2; Standard probability | 64.1% | 67.6% | 65.8% |
|  | *MAX*=2; Contextualized probability | N/A | N/A | N/A |
| (*STD, Disease*)-sanitization | *MAX*=1; Standard probability | 73.3% | 29.7% | 42.3% |
|  | *MAX*=1; Contextualized probability | 73.3% | 59.4% | 65.6% |
|  | *MAX*=2; Standard probability | 62.1% | 97.3% | 75.8% |
|  | *MAX*=2; Contextualized probability | 61.2% | 100% | 75.9% |

Several conclusions arise from the analysis of these results. The best results from the privacy protection perspective (i.e., recall) are achieved when groups of terms are considered in the analysis (i.e., MAX=2), which produces nearly perfect results in all cases. Obviously, this implies evaluating a larger number of term combinations, which also requires a significantly larger number of queries to a WSE and, hence, increases the time needed to sanitize the document. The run time aspect will be discussed latter in this section. However, it is important to note that a cardinality of 2 seems to be enough to detect almost all the risky terms and that further analyses of larger combinations are not really necessary. On the other hand, this setting suffers from a lower precision, which may hamper the utility of the output due to the larger number of unnecessarily sanitized terms. The low precision is caused, in general, because probability calculations suffer from data sparseness when dealing with queries with a large number of terms [26].

On the other hand, the effect of the contextualized probability calculation for groups of terms over the precision is less noticeable in comparison with the analysis of individual terms. Indeed, the fact that we are evaluating groups of terms and, thus, querying them together to the web search engine already minimizes term ambiguity. Moreover, precision slightly decreases as the generalizations used as thresholds in the model instantiation become less specific because i) most of the sensitive terms appearing in the document have been already detected and ii) the stricter analysis tends to increase the number of false positives.

In contrast, if we limit the analysis to individual terms (i.e., MAX=1), recall lowers significantly because combinations of individually innocuous terms (which are risky in aggregate) remain unprotected. In fact, for *C-sanitization*, for which no generalizations are used as thresholds (i.e., *HIV*-sanitization and *STD*- sanitization), and with the standard probability calculation, which is hampered by language ambiguity, (see row 1 of tables 1 and 2) the recall is so low that the output will hardly avoid disclosure (i.e., 23.8% for HIV and 18.9% for STD). In this case, (*C, g(C)*)-*sanitization* helps to make the sanitization process stricter, thus increasing the recall. The same can be said when generalizations are used to contextualize the probability calculation proposed in section 4 because, since we are considering individual terms (which can be potentially ambiguous), the inclusion of the generalizations in the queries helps to make the probabilities more precise, while ensuring the numerical consistency with respect to the taxonomic subsumption. As a result, recall significantly increases as generalizations become less specific (i.e., 76.2% for (*HIV*, *Virus*)-sanitization, 95.2% for (*HIV*, *Infection*)-sanitization and 59.4% for (*STD*, *Disease*)-sanitization). Note also that contextualized term probabilities can only be applied to the (*C, g(C)*)-*sanitization* model, since we need the generalization defined as sanitization threshold to contextualize the queries.

The results of the analysis suggest that the best configuration from the privacy protection perspective is such in which groups of terms with cardinality around 2 are considered (i.e., highest recall). However, a better balance between utility and protection can be achieved in the case of analyzing individual terms but using an adequate generalization as threshold and contextualizing term probabilities. In any case, an additional dimension should be considered in order to prefer one approach or the other: the run time of the sanitization process. In Table 3 we provide the average evaluation results (i.e., precision, recall and F-measure) of the different model instantiations and documents (i.e., *HIV* and *STD*) showed in the previous tests (entities and model instantiations) together with the average run time required to sanitize the documents for each parameter value.

Table 3. Average precision, recall, F-measure and run time for the different implementation parameters.

| *Implementation parameters* | *Avg. Precision* | *Avg. Recall* | *Avg. F-measure* | *Avg. time* |
|---|---|---|---|---|
| *MAX*=1; Standard probability | 89.9% | 28,8% | 43% | 1 min. |
| *MAX*=1; Contextualized probability | 77.2% | 76.9% | 76.3% | 2 min. |
| *MAX*=2; Standard probability | 61.4% | 91.1% | 72.7% | 40 min. |
| *MAX*=2; Contextualized probability | 62.4% | 100% | 76.8% | 41 min. |

Run time differences are very noticeable when moving from the analysis of individual terms to combinations of just two terms. Indeed, in the former scenario, the computational cost scales linearly with respect to the number of terms in the document, whereas in the latter the cost is exponential with respect to the number of terms and the cardinality of the groups. Even though the actual cost of evaluating groups of terms is minimized in practice thanks to the heuristics implemented in the proposed Algorithm 1, the analysis of term combinations incurs in a high cost because of the large number of queries that are required. This may be problematic for large documents or contexts. In such cases, the highly scalable analysis of individual terms may be preferable, especially under the consideration that the sanitization can be made stricter (i.e., to increase the recall) by using less specific generalizations as thresholds (e.g., the output of (*HIV, infection*)-sanitization achieves a recall of 95,2% in around 2 minutes).

On the other hand, the average evaluation measures shown in Table 3 suggest that the contextualization of probabilities we propose in section 5 adds no significant execution overhead over the analysis (since the number of queries remains the same, but just involving more terms), and also helps to improve the results in all cases. It is important to note that the increase in run time resulting from the contextualization of individual terms (1 vs. 2 minutes, in average) can be explained by the larger number of sanitizations resulting from the more precise analysis, which requires additional queries to assess and select the generalization to be used as replacement of risky terms.

## 6. Conclusions and future work

The main goal of our work is to offer a theoretically sound and practically feasible solution to assist the burdensome task of manual document sanitization and to do so by mimicking the semantic reasoning employed by human sanitizers. To advance in that direction, in this paper we presented the theory and developed the practical aspects of (*C, g(C)*)-*sanitization*, a flexible and inherently semantic privacy model for plain textual documents that, in comparison with other works focusing on documents sanitization [15, 22, 23], offers an intuitive instantiation and clear beforehand privacy guarantees to practitioners.

The main contributions of our work are: i) we proposed a flexible model that allows to intuitively configure the trade-off between the desired level of protection and of utility preservation; ii) we designed a customizable and scalable algorithm implementing the model that is driven by several semantics-preserving heuristics; iii) we detailed a general solution to coherently and accurately compute the term probabilities needed by the information theoretic enforcement of the model by using the Web and web search engines as proxies for social knowledge. As a result, we provided the tools to ensure the applicability and accuracy of (*C, g(C)*)-*sanitization* in practical scenarios, putting special emphasis on the preservation of data utility and in ensuring the scalability of the implementation. Empirical results and comparisons against the baseline showed that such tools contributed to improve the accuracy of the actual protection and also provided the flexibility to tune the behavior of the implementation according to specific needs (privacy, privacy/utility ratio and/or scalability).

As future work, in order to further illustrate the convenience and benefits of our model, we plan to engineer a set of case studies that show: i) how (*C, g(C)*)-*sanitization* can be instantiated in accordance with the legislations on data privacy that are available for different areas (healthcare, finances, census, etc.) and, ii) how it can be applied to a variety of textual inputs (e.g., unstructured medical records, messages to be published in social media, e-mails, documents to be declassified, etc.). Moreover, we also plan to extend the implementation to support a variety of knowledge bases (i.e., other domain specific KBs such as SNOMED-CT or MeSH for medical data) and languages (by relying on multi-language linguistic tools to extract terms, such as OpenNLP and NLTK, and the translations available for the KBs to retrieve suitable generalizations).

## Acknowledgements and disclaimer

This work was partly supported by the European Commission (projects H2020-644024 "CLARUS" and H2020-700540 "CANVAS"), by the Spanish Government (projects TIN2014-57364-C2-R "SmartGlacis", TIN2015-70054-REDC "Red de excelencia Consolider ARES" and TIN2016-80250-R "Sec-MCloud") and by the Government of Catalonia under grant 2014 SGR 537. The opinions expressed in this paper are those of the authors and do not necessarily reflect the views of UNESCO. M. Batet is supported by a Postdoctoral grant from Ministry of Economy and Competitiveness (MINECO) (FPDI-2013-16589).